# Potential of EPUB3 for Digital Textbooks in Higher Education

**Martin Ebner**
*Graz University of Technology, Graz, Austria*

**Christian Gailer**
*Graz University of Technology, Graz, Austria*

**Mohammad Khalil**
*Graz University of Technology, Graz, Austria*

**Michael Kopp**
*University of Graz, Graz, Austria*

**Elke Lackner**
*University of Graz, Graz, Austria*

**Michael Raunig**
*University of Graz, Graz, Austria*

**Abstract**

*The e-book market is currently in a strong upswing. This research study deals with the question which practical uses the e-book format EPUB3 offers for (higher) education. By means of a didactic content analysis, a range of interactive exercise types were developed as a result of conversations with teachers. For this purpose, a didactic and technical concept has been developed. Different kinds of exercises were prototypically implemented in an e-book. Finally, a brief overview reflects the present state of the current e-book readers.*

*A subsequent discussion illuminates the strengths and weaknesses of the format. In summary, it can be remarked that EPUB3 is suitable for a variety of different exercises and that it is able to serve as a basic format for forthcoming digital textbooks. Furthermore the openness of EPUB3 will assist Open Learning and Teaching in a meaningful way.*

**Introduction**

These days, electronic books (e-books) benefit from a great distribution in the fields of poetry and fiction. The market share in the Anglo American area has just recently reached 20% (Wischenbart, 2013). Compared to that, Europe lies on the lower end – but there is a great potential for growth. In Germany for example, there is an annual growth rate of 60%[7] in the field of e-books. Since Amazon has launched its e-book reader Kindle in 2007 and Apple its iPad in 2010 e-books are becoming more and more popular. Due to the fact of the growing number of devices the topic of digitalized books has experienced its renaissance (looking back on a long history starting in the early 1970s with the "Project Gutenberg" [8]). Only a couple of years ago the number of bought e-books was lower than 1% (Ehling, 2011). On the other side today's surveys among university freshmen point out that one out of seven students already own their own e-reader (Ebner et al., 2014).

However, high quality content designed in an appropriate and didactically meaningful way is a precondition for effective learning as well as for teaching. Within university education the printed book has already been completed by digital scripts e.g. using MS Word, Adobe PDF, LaTeX or other text based formats. Those documents are widely distributed via Learning Management Systems or different platforms. So it can be stated that the hardcopy is becoming more and more a minority, replaced by digital documents (Nagler et al., 2011). Especially the flexible character of those documents and the easy way to handle them with (mobile) devices gather the attention of today's learners.

Nevertheless, the definition of an e-book is not easy due to the fact that in the early days of digital books each single digitalized document represented by the format PDF has been called "electronic book". Today there are more or less three meanings of that term:

- E-book = the digital book: As described before, an e-book can be just a digital copy of a former printed book. This definition includes not only the digital version of real books, but also journals, papers, magazines or even lecture handouts.

- E-book done with authoring tools: In the field of technology enhanced learning software for doing HMTL-based content was introduced already 10 years ago. With the help of such tools, like the ABC Publisher[9], lecture content is created, published and used in an educational context.

- E-book for e-readers: This means books especially designed for e-readers and Tablet computers. In general, such devices are able to display traditional PDF, but also different formats like EPUB, which allow displaying the content in a much more flexible way.

In our publication we are strongly interested in the open standard format EPUB. Due to the fact that Open Learning and Teaching are strongly based upon open educational resources our research concentrates on open standards such as EPUB. The EPUB (electronic publication) format is an e-book standard by the International Digital Publishing Forum (IDPF) and was firstly launched in 2007. The IPDF published the EPUB3 standard in 2011 (Conboy et al., 2011). This new format opens up numerous opportunities for teaching by utilizing state-of-the-art technology.

EPUB3 is a container format. A "container" is represented by a simple ZIP file containing all relevant publication data. Thus, EPUB3 is essentially based on current web technologies like

---

[7] Mediacontrol: eBook-Trend 2013: http://www.ceebo.de/news/jahrescharts2013.html [March 2014]
[8] http://www.gutenberg.org/ [March 2014]
[9] https://ebook.tugraz.at/publisher/ [March 2014]

HTML5, CSS3 and JavaScript. Embedding multimedia formats such as audio, video or vector-based images as well as custom fonts and mathematical formulas is possible. Furthermore, the format supports the text-to-speech functionality and thereby computer-assisted reading of text content.

**Research Questions and Goals**

Due to the fact that previous e-book formats did not offer many additional benefits in comparison to traditional textbooks, e-books have been largely ignored in education so far. Thanks to the growing audience for digital e-readers (Südwest, 2012), such as tablet computers and e-ink devices, and the resulting widespread availability of digital media, e-books may play a more important role in the near future. Thus, in the near future digital textbooks may play a significant role in education (Ebner et al., 2013).

The question of which role e-books may take in education and how they can usefully supplement tuition has been examined in a thesis by Monika König (König, 2013). According to that from a theoretical point of view there is nothing hindering the introduction of e-books apart from the question of financing.

Currently, the prevalence of the format is undertaking a transition from EPUB2 to EPUB3 (Wenk, 2013). However, the advantages of the newer format outweigh those of the previous version, and therefore EPUB3 is continuously finding more distribution. This process takes time and depends on a number of factors. Enhancements are needed in the entire publication process (Bläsi et al., 2013), such as the distribution by publishers and the user's acceptance.

Nevertheless, this transition is constantly taking place, and therefore the EPUB3 format should be subjected to a more accurate examination. The continuous progress raises the following questions:

- What are the potentials of the EPUB3 format?
- Which contents are suitable for the format?
- What are the strengths and weaknesses of the format?
- What should the didactical and technical preparation process look like considering the field of (higher) education?

We try to answer these questions with the help of a field study and by developing a French textbook. Thereby, common work sheets and exercises have to be implemented into the e-book to point out the strengths and weaknesses of the format in a technical and a content-related sense. Using a didactical content analysis and cooperating with teachers, we managed to cover a wide array of possible contents and implement them by means of a prototype.

**Related work**

In one research study (Fenwick et al., 2013) interactive exercises are integrated in an e-book. It is a textbook for the programming language "Prolog". Using the authoring tool Apple iBooks Author[10], a prototype of a textbook in iBooks format[11] is developed. The focus of the interactive elements is on image galleries und multiple choice exercises which are already included by default in iBooks Author. The prototype was evaluated by a group of 18 students. In another research study (Singhose et al., 2013) an existing textbook in the field of mechanics was implemented in an iBooks format. For the presentation of complex mechanical processes, interactive elements like image galleries and animations were used.

---

[10]    http://www.apple.com/at/ibooks-author/ [March 2014]
[11]    https://itunes.apple.com/de/app/ibooks/id364709193?mt=8 [March2014]

In a third research work (Gavrilis et al., 2013) an e-book in EPUB format was developed. The user input is forwarded to a server via client-server communication. A small reader was developed for mobile devices with the Android operating system. EPUB3 publications, however, lacked Java libraries for reading, so the older EPUB2 format was used.

**Content Analysis**
Considering the improvements the EBUB3 format is accompanied by, e-books are becoming more attractive for education purposes. The efficiency of teaching and studying processes can be greatly improved by the use of methods of instructional design such as integration of interactivity or the implementation of multimedia (Kerres, 2007). It should be noted however, that using these technical abilities does not automatically lead to an increase of quality of the teaching. This requires developing a dedicated didactic or instructional design (Baumgartner et al., 2013).

**Didactical concept**
Talking to language teachers, we conducted a didactical content analysis of the teaching contents which should take into consideration a) the different learning styles and types, b) the different levels of knowledge and c) the four communicative language competences stated by the Common European Framework of Reference for Languages (CEFR)[12]. Learning a language means to learn, to understand (i.e. by listening and reading), to speak (i.e. by spoken interaction and spoken production) and to write. A good textbook to foster language learning should regard to these four competencies (in different ways). Furthermore, it is necessary to bear in mind that in university language courses the audience is likely to be heterogeneous. Students do these courses for at least one of these two reasons: a) to refresh their language skills, or b) to learn a new language. The exercises should not be only drill and practice exercises in a behavioristic way (Morgret et al., 2001), but should be varied, appealing, interactive, motivating and multimedia-based. Last but not least, the exercises should be auto corrective in order to assure immediate feedback. The learners do an exercise and immediately check their results. They are able to learn or to train themselves autonomously and in a self-determined way regarding to their own learning rhythm and needs (Chardaloupa et al., 2013).

At first, we defined the three main categories for learning French, besides learning the vocabulary: a) phonetics, b) grammar and c) conjugation. We addressed issues regarding appropriate exercise examples for each of those categories. These exercise examples should activate different sensory channels (e.g. visual, acoustic) and different learning levels (e.g. beginners, specialists). The exercises were again divided and assorted into sub-categories. To give an example: When talking about phonetics it is important to a) recognize a phoneme before b) pronouncing it in the right way. Therefore it is necessary to design exercises where the learners have to differentiate between a [p] and a [b] before reading words aloud on their own. When they finally come to language production feedback would be nice. So the learners read words, phrases or texts and can check the solution where a native repeats the same words, phrases or texts in a slow, a normal or a fast way. Even if this exercise type is not self-checking, the students can repeat the exercise and train their receptive and productive skills and competences. If the words, the phrases or texts are also written down, the learners can read them simultaneously – more than one sensory channel will be activated. At a further step, the learners will hear phonemes, words, phrases or texts and will have to write them down. The checking of this exercise type can be in form of a sample solution that pops up after finishing the production part.

---

[12] http://www.coe.int/t/dg4/linguistic/Cadre1_en.asp [March 2014]

The same holds for grammar exercises. The learners should start by recognizing a verb's right inflexion before trying to conjugate a verb on their own. Dividing these to exercises levels, language students will be able to train their understanding and their productive skills. Beginners will probably have to start by the recognizing part whereas experts or refreshers can directly go to do the production part. As in French diacritic signs (e.g. accents) are indispensable, they should be easy to enter. Gap-filling exercises should be available in two versions, as real gap filling exercises and as a solution grabbing exercise, where the solutions are displayed in a dropdown list.

Last but not least, we stressed out the importance of images and animations for language learning (e.g. for motivating and creating appealing exercises and no plain text exercises). Images and animations are not only interesting impulses to start conversations or text production, they can also illustrate grammar chapters (e.g. the well-known "house"[13] for the French passé compose tense). There should therefore be exercises that combine images and text or audio input. This exercise type is well known when learning vocabulary as, for example, the parts of the body, but it can also be used to learn prepositions or other grammatical aspects. By choosing specific images the same exercise can be adapted for children and for adult learners – as we think of vocabulary work concerning fruits for example. Not every exercise type is appropriate for young as well as for adult learners. The crossword for example seems to need some experience in doing crosswords and is not as self-explanatory as a memory or a drag-a-text-and-drop-it-on-an-image exercise known by children from the "real world".

As a result of these reflections and needs, a list of different exercise types was derived and will be presented in part 4 of this paper.

**Technical concept**
JavaScript was employed as the programming language to implement interactive exercises and contributes to interactivity in e-books. In addition, it can be used to analyze the user input and is of high importance for both creating and changing content.

The core of the system is patterned according to the Model-View-Controller (MVC) concept. **Fig. 14** shows a graphical illustration of the MVC concept. The tasks of the three units are explained in more detail here. The **model** unit contains the content of the exercise, such as tasks, possible answers and solutions, options for multilingualism as well as any legal licensing information for multimedia contents. Furthermore, user input is stored in the model. The **view** unit is responsible for the presentation of the content. Interactive exercises are created dynamically by

---

[13]http://www.cliffsnotes.com/foreign-languages/french/french-i/french-i-the-passe-compose/the-passe-compose-with-etre [March 2014]

DOM (Document Object Model) manipulation based on the model data at runtime. Furthermore, there are methods for creating popup and dialog windows. Certain additional features, such as image magnification or adding copyright information to multimedia content can be activated via parameter. The **controller** unit initiates the dynamic creation of the exercises, checks user input and creates evaluations. With the evaluation of the epubReadingSystem object, a differentiation occurs with respect to the reading system used.

**Fig. 14.** Model-View-Controller concept

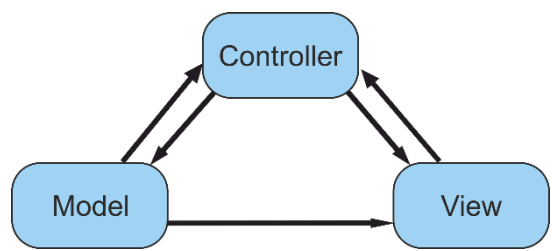

For implementation, the two JavaScript libraries jQuery Mobile[14] and Modernizr[15] are used. jQuery is a widely used free library and is particularly suitable for the selection and manipulation of DOM elements. Modernizr is used to verify the e-reader and its rendering engine on functionality.

**Implementation of a Prototype**
One goal of this research work was the prototypical implementation of derived exercise types. All of the exercise types were implemented into the e-book based on a model example. Thanks to HTML5 and JavaScript support, a high level of interactivity is possible.

The implementation was carried out by taking into account the simplest possible way of reusability. This is a first step toward the automatic generation of e-books with interactive exercises. This form of e-book production could be realized with the help of widget libraries. So, the authors will not need to have any programming skills, thereby increasing production speed and the number of authors significantly while costs could be kept low (Sigarchian et al., 2013).

This form of reuse requires a strict separation of practice content and graphical display. The contents are put into separate files within the EPUB container. The practice content is defined using object literals. In **Fig. 15**, such a construct can be seen by the example of a multiple choice type exercise. This exercise, for example, can be inserted at that position of the e-book where a *div* element with the ID *mc1* is available. Exercise content can be exchanged as desired without having to deal with the routine programmed in the view unit. An automatic generation of these files is subsequently conceivable, for example, via online forms.

---

[14] http://jquerymobile.com/ [March 2014]
[15] http://modernizr.com/ [March 2014]

**Fig. 15.** Nested object literal with content of the exercise type "Multiple-Choice"

```
1   // nested object literal
2   var multiplechoiceInput = {
3       // question: question phrase
4       // answer: given answers separated by a semicolon
5       // correctAnswers: correct answers indices separated by a semicolon
6       // multiSelect: can be "true" or "false", one or all are selectable
7       // multiMedia (optional): type: image (.jpg), audio (.ogg AND .mp3), video (.mp4 AND .webm)
8       //                        file: filename without extension
9
10      // this must be same as id of <div class="multiplechoice" id="mc1" /> element in xhtml file
11      mc1: {
12          task1: {
13              question: "De quelle couleur est cette fleur?",
14              answers: "bleu;pourpre;jaune",
15              correctAnswers: "2",
16              multiSelect: "false",
17              multiMedia: {
18                  type: "video",
19                  file: "butterfly"
20              }
21          },
22          task2: {
23              question: "Quelles langues sont parlées dans ce poème."
```

To evaluate the different forms of representation the prototype was created with both, a fixed and a dynamic layout.

Considering the didactical concept and the defined categories as well as technical possibilities different exercises have been created:

─ Drag and Drop
- Pair Assignment: Coherent elements are connected via drag and drop. A possible implementation can be seen in **Fig. 16**. The screenshots have been made using the e-reader Readium[16].
- Group Assignment: The user has to move elements to their appropriate groups, e.g. allocating articles to nouns.
- Order Assignment: Sorting elements in the correct order.
- Drag and Drop on Image: Dragging elements to the right place on an image.

─ Text Assignments
- Cloze: Filling out the cloze text by entering the correct words.
- Dictation: Playback of audio files and entering text at the same time for the means of determining orthographical skills.

─ Quizzes
- Multiple-Choice: There are several possible answers to choose from. After successful verification, the user gets to the next question. A screenshot of the exercise is shown in **Fig. 17**.
- Text Quiz: Answering questions by entering the text.
- Crossword Puzzle: Filling out a crossword puzzle and subsequent verification of the solution word.

─ Selection Tasks
- Drop-Down List: Selection of possible answers out of a list.
- Memory: Finding matching pairs by flipping cards. A screenshot of this exercise is shown in **Error! Reference source not found.**.
- Text Selection: Selecting single words or text phrases.

---

[16] http://readium.org/ [March 2014]

─ Media-Overlay: This means synchronizing text and recorded voice files. While playing the voice files, the related text passages are highlighted.

**Fig. 16.** Screenshot of the exercise "Pair Assignment" (Reader: Readium)

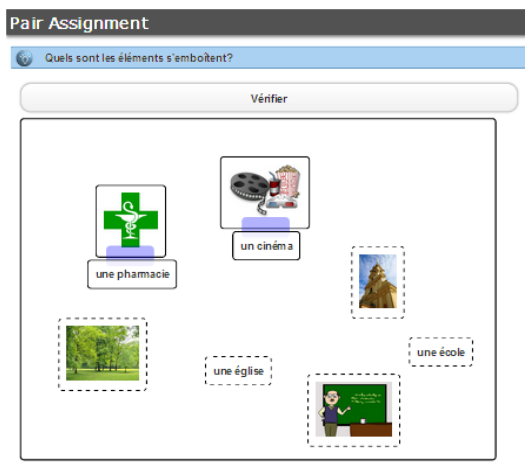

**Fig. 17.** Screenshot of the exercise "Multiple-Choice" (Reader: Readium)

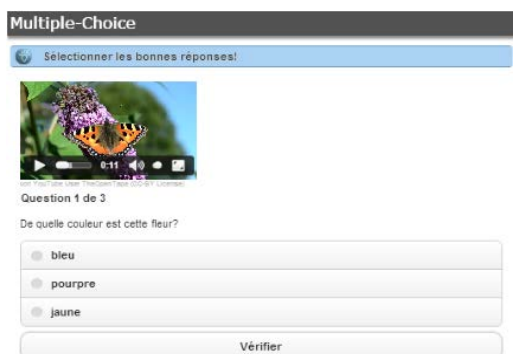

**Evaluation**

As we presented the prototype to a wider audience (of language learners, teachers and non-language learners) they were at first sight impressed by the high level of interactivity and multimedia implementation made possible by EPUB3. At the second sight especially the teachers were delighted by the appealing layout and the fact that the exercises covered the different language competences and different proficiency levels as stated by the CEFR. The exercises activate more than one sensory channel and use images that are licensed as creative commons, so the e-book could be seen as an open educational resource. The license information was clearly displayed. One person stressed out the point that the exercises can be done not only by adults but also by children and mentioned the pair assignment and the memory.

**Fig. 18.** Screenshot of the exercise "Memory" (Reader: Readium)

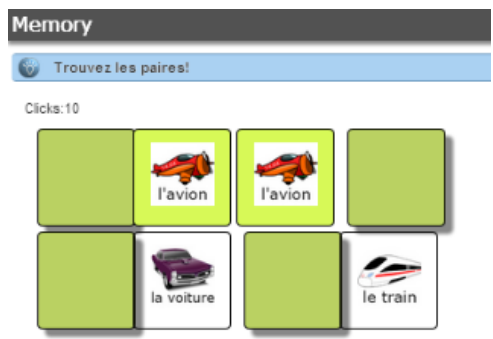

It is for sure that there will always be additional wishes and needs that the audience expresses as for example regarding the lack of annotation possibilities or the lack of advanced correction modes for writing exercises besides the sample solution given in the prototype. Further feedback concerned the integration of videos and their annotation. It would be desirable to have the possibility to integrate not only videos but videos including questions that have to be answered in order to continue watching the video. It would also be nice, according to the given feedback, to have the possibility to annotate the video or to share annotations with colleagues. By this means, the social learning component could be implemented.

For the language learners, it was important that the e-book is not limited to Apple devices (as, for example, iBooks are) and that the exercises were self-explanatory and easy to handle. There were wishes expressed by learners: to be able to save the results and to continue their training process at a certain point and to communicate with each other or the language teachers. From their point of view, collaborative or interactive exercises would therefore be as desirable as the communicative aspect is. The language learners are often used to communicate and interact with each other in social networks such as groups on Facebook or circles on Google+. If the e-book could integrate communication possibilities such as learning groups or at least sending e-mails to each other, the learners would use them, as they assured.

**Lessons Learned**
The EPUB3 allows the representation of differentiated content for learning and teaching. It can be used for teaching new content as well as for reviewing already learned material. Knowledge can be strengthened through interaction and active engagement with learning content. In addition, a high motivation potential is reached. With the development of its own user interface, content can be individually organized and structured by the authors.

It has been shown that in principle all content such as text, interactive exercises, animations and multimedia content can be presented and structured in similar ways, as modern web technologies allow.

Another advantage is the possible network abilities. For example, a promotion of social skills through knowledge sharing is as well conceivable as the monitoring of the learning progress on the basis of client-server communication.

Some insightful points could be observed during the implementation of the prototype: It was found that the integration of interactive exercises is only suitable for e-books with a fixed layout. The fixed layout enables the exact positioning of the exercise within a page. E-readers, however, do not differ in dynamic layouts, whether it is regular text or a self-contained exercise section. Thereby it is possible, that a page break is set amidst the exercise area and carrying out the exercise is no longer possible. One approach to solve this problem is the manual opening of interactive exercises within distinct dialog boxes. Furthermore, a combination of fixed and dynamic layouts is feasible.

It has been observed that teachers often desire animation features. Animations can make an important contribution to the learning process and serve to enliven the content. While HTML5 animations are certainly possible, the development of customized animations demands quite a high level of development effort depending on their complexity. Furthermore, it also requires reconciliation to the limited performance of mobile reading devices.

Furthermore, some obstacles were observed. Due to the usually small screen size of e-reader devices there is a need for a well-planned layout and structuring of the content.

With drag and drop exercises, in which one drags an element from one location to another, the dragging movements are often misinterpreted as swiping gestures. This means that unwanted pages are turned.

The integration of multimedia content may lead to a high memory consumption of the EPUB containers. In addition, multimedia files should be stored in multiple formats in order to increase compatibility. This redundancy results causes a further increase of the file size. Therefore, the advantages and disadvantages of local and online stored multimedia contents should be considered.

Similar to different web browsers and their rendering engines, there are partially different interpretations of standards with EPUB3 readers. An examination of the reading system used and an appropriate reaction at runtime is very important especially in highly interactive exercises.

A device testing of the interactive exercises was carried out with the following tablet computers:

- Asus Eee Pad Transformer Prime (Model-No.: TF201-1I066A, Android 4.1.1)
- Samsung Galaxy Tab 7" (Model-No.: GT-P1000, Android 2.3.6)
- iPad 3 (iOS 7.0)

These devices have no factory-fitted software for reading e-books. However, such software can be installed. To this end, a search of available e-readers was made before the test. For Android devices, the Himawari Reader[17] was used, as it showed the best results at that time. For the iPad, the e-reader iBooks was used. In Table 1 a list of those results can be seen.

One can say that the interactive exercises on Android devices with version 4.1 as well as in Apple iBooks are working fine. Nevertheless it should be noted that Media Overlay in iBooks works only with e-books with a fixed layout. On the device with an older version of Android, many of the exercises could not be performed.

---

[17] https://play.google.com/store/apps/details?id=jp.green_fld.himawari [March 2014]

In Table 1, the duration which was spent on the implementation of the particular exercise is also mentioned. These details refer to the programming of the exercises and allow an approximate estimate of the costs. The conducted content analysis, setting up the structure of the EPUB container and any additional functions are not included in these values.

**Table 1.** Duration of implementation and device test

|  | Asus Eee Pad | Samsung Galaxy | Apple iPad 3 | Duration [h] |
|---|---|---|---|---|
| **Pair Assignment** | ✓ | ⊖ | ✓ | 10 |
| **Group Assignment** | ✓ | ⊖ | ✓ | 8 |
| **Order Assignment** | ✓ | ⊖ | ✓ | 8 |
| **Drag&Drop Image** | ✓ | ⊖ | ✓ | 6 |
| **Cloze** | ✓ | ⊖ | ✓ | 2 |
| **Dictation** | ✓ | ⊖ | ✓ | 2 |
| **Multiple-Choice** | ✓ | ⊖ | ✓ | 12 |
| **Text Quiz** | ✓ | ⊖ | ✓ | 2 |
| **Crossword Puzzle** | ✓ | ⊖ | ✓ | 5 |
| **Drop-Down List** | ✓ | ✓ | ✓ | 8 |
| **Memory** | ✓ | ⊖ | ✓ | 6 |
| **Text Selection** | ✓ | ⊖ | ✓ | 4 |
| **Media Overlay** | ✓ | ⊖ | ✓ | 5 |

**Conclusion**

EPUB3 has the potential to play a significant role in supporting the teaching and learning process and to make a significant contribution in the field of Open Learning and Teaching. This research shows that a wide range of content can be implemented in EPUB3 format. In discussions with teachers a wide variety of content and possible exercises was developed. The amount of content plays a subordinate role. An e-book can deal with a small specific subject as well as extensive materials over several chapters.

In addition, the strengths and weaknesses of the format in relation to digital textbooks were identified. The strength of the format was found to be the high level of interactivity with the user. This results in a high level of motivation, which could be demonstrated by implementing a prototype. Since EPUB3 is an open format, this has a positive impact on its spread. Even now it is supported by platforms and devices from different manufacturers, although not yet fully.

The circumstance of the open format is an advantage as well as a disadvantage. The standard describes only the requirements of the format of the reading system, but not the implementation. While publishers of proprietary formats provide corresponding reading software, the reading software of open formats is produced by third party developers. However, these developers often implement only parts of the standard which results in quality differences between reading software.

Crucial for the learning success is a well-planned selection and structuring of the content. During this process, the benefit through the strategic use of media didactics compared to conventional textbooks should be taken into account. It could be shown that the format is suitable for both children's textbooks as well as textbooks in higher education.

According to the current state, extensive knowledge in the areas of HTML5, CSS3 and JavaScript, as well as EPUB specifications are necessary for the creation of EPUB3 publications. Decisive for the implementation of large amounts of teaching materials is primarily the presence of easy-to-use authoring tools.

This topic is meant to be further developed in future work. In particular, the implementation of a field study for a practical evaluation of the prototype by students is desirable.